# Neighbourhood topology unveils pathological hubs in the brain networks of epilepsy-surgery patients

**Leonardo Di Gaetano,[1] Fernando A. N. Santos,[2] Federico Battiston,[1] Ginestra Bianconi,[3,4] Nicolò Defenu,[5] Ida A. Nissen,[6] Elisabeth C. W. van Straaten,[6,7] Arjan Hillebrand[6,8,9] and Ana P. Millán[6,10]**

Pathological hubs in the brain networks of epilepsy patients are hypothesized to drive seizure generation and propagation. In epilepsy-surgery patients, these hubs have traditionally been associated with the resection area (RA): the region removed during the surgery with the goal of stopping the seizures, and which is typically used as a proxy for the epileptogenic zone. However, recent studies hypothesize that pathological hubs may extend to the vicinity of the RA, potentially complicating post-surgical seizure control. Here we propose a neighbourhood-based analysis of brain organization to investigate this hypothesis. We exploit a large dataset of pre-surgical magnetoencephalography-derived whole-brain networks from 91 epilepsy-surgery patients. Our neighbourhood focus is 2-fold. Firstly, we propose a partition of the brain regions into three sets, namely resected nodes, their neighbours and the remaining network nodes. Secondly, we introduce generalized centrality metrics that describe the neighbourhood of each node, providing a regional measure of hubness. Our analyses reveal that both the RA and its neighbourhood present large hub status, but with significant variability across patients. For some, hubs appear in the RA; for others, in its neighbourhood. Moreover, this variability does not correlate with surgical outcome. These results highlight the potential of neighbourhood-based analyses to uncover novel insights into brain connectivity in brain pathologies, and the need for individualized studies, with large enough cohorts, that account for patient-specific variability.

1 Department of Network and Data Science, Central European University, Vienna 1100, Austria
2 Institute for Advanced Study, University of Amsterdam, Amsterdam 1012 GC, The Netherlands
3 School of Mathematical Sciences, Queen Mary University of London, London E1 4NS, UK
4 The Alan Turing Institute, British Library, London NW1 2DB, UK
5 Institute for Theoretical Physics, ETH Zurich, Zurich 8093, Switzerland
6 Department of Clinical Neurophysiology and MEG Center, Amsterdam UMC, Vrije Universiteit Amsterdam, 1081HV Amsterdam, The Netherlands
7 Academic Center for Epileptology Kempenhaeghe and MUMC+, Heeze 5591 VE, The Netherlands
8 Amsterdam Neuroscience, Brain Imaging, 1081HV Amsterdam, The Netherlands
9 Amsterdam Neuroscience, Systems & Network Neuroscience, 1081HV Amsterdam, The Netherlands
10 Department of Electromagnetism and Matter Physics, and Institute 'Carlos I' of Statistical and Computational Physics, University of Granada, 18071 Granada, Spain

Correspondence to: Ana P. Millán
Department of Electromagnetism and Matter Physics
Av. Fuentenueva s/n, Granada 18071, Spain
E-mail: apmillan@ugr.es

**Keywords:** epilepsy; brain networks; magnetoencephalography; brain hubs; topological data analysis

**Graphical Abstract**

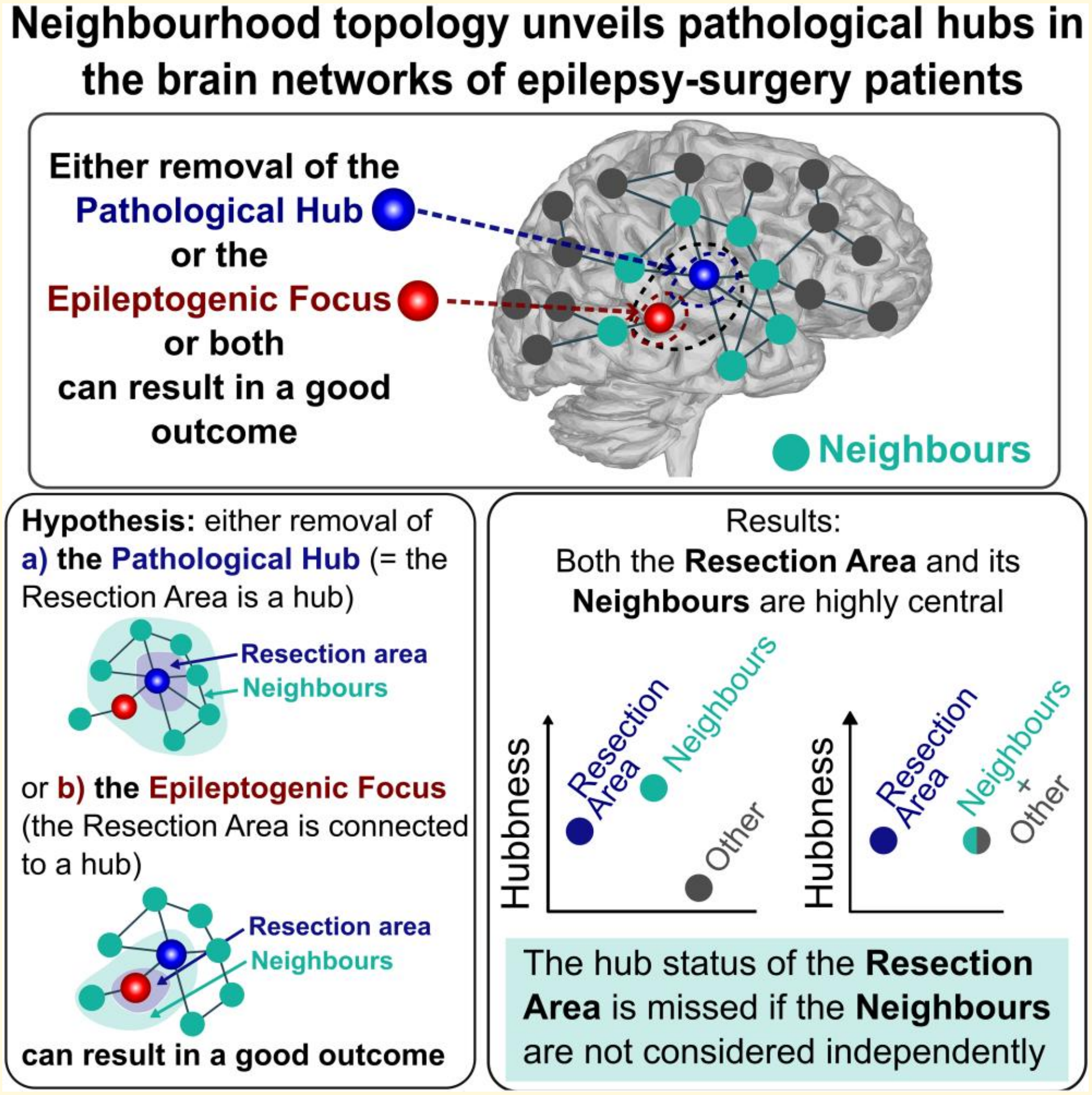

# Introduction

The network description of complex systems, such as the brain, provides a remarkable tool to unveil its underlying organization and emergent dynamics. Such a description has enriched our understanding of brain organization both at the macroscopic[1-4] and microscopic[5] levels and has found remarkable clinical applications.[6,7] A notable example, which is the focus of this study, is the case of epilepsy surgery. This is the recommended treatment for drug-resistant epilepsy patients, and it entails the removal or disconnection of a set of brain regions—the epileptogenic zone (EZ)—with the goal of stopping seizure generation and propagation.[8-10] In practice, there is no gold standard to identify the actual EZ; instead, the EZ may be approximated by the resection area (RA) in combination with surgical outcome: for patients with a good outcome, the EZ is included in the RA, whereas for patients with a bad outcome, the EZ was at least partially preserved by the surgery. Epilepsy surgery is preceded by an extensive pre-surgical evaluation, involving different imaging modalities such as MRI or electro- and magneto-encephalography (E/MEG). However, positive outcome rates (i.e. seizure freedom after the surgery) are not optimal, and around 30% of the patients continue to present seizures 1 year after the resection, although this number can go up to 50% for cohorts with complicated aetiology.[11] With the goal of improving these outcome rates, network-based studies have investigated in detail the brain network organization of epilepsy patients to unveil pathological effects that may predict surgical outcome.[12,13]

Within this context, a big conceptual leap has taken place, from the notion of individual EZs to the consideration of epileptogenic networks that arise from the interplay between different brain regions in promoting and inhibiting ictal activity.[14-17] According to this perspective, the effect of a given surgery cannot be determined alone by the characteristics of one or more regions but needs to be measured against the whole epileptogenic network.[18] Data-driven and modelling studies seem to support this hypothesis, and thus, network mechanisms are recognized to participate in the generation and propagation of seizures.[19-28]

Substantial evidence underscores changes in structural and functional brain networks in epilepsy,[29,30] particularly related to the EZ.[31-33] Whether there is an increase or decrease in connectivity of the EZ compared to healthy individuals, however, remains an open question. fMRI-based studies initially pointed towards a disconnection of the EZ,[34-37] which has been supported by some invasive EEG[38] and MEG studies.[39-41] However, recently several M/EEG studies have suggested hyperconnectivity of the EZ and neighbouring regions.[15,31,42-48]

Pathological changes in brain connectivity in epilepsy are disproportionally associated with the network hubs[49]—highly central or important regions in the network architecture of the brain—a finding echoed in other neurophysiological disorders such as Alzheimer's disease, multiple sclerosis or stroke.[29,50] In the case of epilepsy, pathological hubs that facilitate seizure generation and propagation may be present. Indeed, the properties of the brain hubs, including their spatial distribution and overlap with the RA, are associated with epilepsy-surgery outcome.[51-56] Notably, however, hubs can also have an inhibitory effect to prevent the ictal state,[38,55] and it should be noted that hub removal can lead to increased side effects from the surgery. The RA and the EZ have been associated with brain hubs by several studies, both in the ictal[57-60] and interictal[45,57,59] states. Such studies found associations between hub removal and seizure freedom with different MEG-based connectivity measures,[45,61] although in a recent study involving a large cohort ($n = 91$) of epilepsy-surgery patients, we could not confirm these findings.[46] In a recent MEG study with a smaller cohort of 31 epilepsy patients, Ramaraju *et al.*[48] were able to classify epilepsy-surgery patients according to surgical outcome (79% accuracy and 65% specificity) by comparing the degree centrality (a measure of *hubness* given by the number of neighbours of a node) of the RA to the remaining network nodes, in the pre-surgical brain-network.

Overall, although hub removal has been associated with a favourable outcome of epilepsy surgery, this does not seem to be a necessary condition for a good outcome. Indeed, brain hubs do not always overlap with the RA, even for seizure-free (SF) patients.[21,22,24,46] These findings motivated the hypothesis that the seizure onset zone (SOZ, the region where seizures start) need not coincide with the pathological hubs but may be strongly connected to them, in which case removal of either the SOZ, the pathological hub, or even the connection between them may be enough to prevent seizure propagation and achieve a good outcome.[21,46] Thus, regional brain organization around the SOZ—as opposed to only its centrality— becomes a promising target for understanding the effect of a given resection.

To study the role of regional brain organization in epilepsy surgery, we propose here a neighbourhood-based description of brain connectivity and of the effect of a resective surgery in epilepsy. Firstly, we explicitly consider the role of the connectivity of the RA neighbours by implementing a partition of the brain regions, namely into RA nodes, their neighbourhood and the remaining network. By doing so, we are able to specifically address the question of the emergence of pathological hubs in the vicinity of the RA, and their relation to surgical outcome. Secondly, we introduce a novel analysis framework to quantify regional brain organization based on the notion of extended neighbourhoods, following a previous theoretical study that generalizes the notion of the clustering coefficient.[62] The extended neighbourhood of a node describes its area of influence, providing a mesoscopic description of brain organization that can inform us of, e.g. the existence of regions with strong recurrent connectivity.[62] By characterizing the extended neighbourhood of each brain region through topological data analysis,[63] we propose the generalization of local node-based centrality metrics to regional descriptors that encode regional organization. As we go on to show, the neighbourhood-based description unveils the distinctive properties of the neighbours of the RA. In this multi-frequency study (including six specific frequency bands as well as the broadband), we found that the RA and its neighbours shared a highly central status that was significantly larger than that of the remaining brain regions. The relative centrality of the RA and its neighbours varied within the population (and between frequency bands and network metrics), and whether the RA or its neighbours were more central was not an indicator of surgical outcome [Area Under the Curve (AUC) = 0.46 for the classification of patient outcome for the broadband network]. In contrast, we achieved a fair classification of the patient groups (AUC = 0.62, 0.64, respectively) when considering either the hub status of the RA or of its neighbourhood in relation to the remaining brain regions. These findings support the notion of the emergence of pathological hubs in the brain of epilepsy patients that may not coincide with the SOZ but appear in its neighbourhood. Overall, our findings highlight the need to consider regional brain connectivity in epilepsy-surgery studies, for instance, with the notion of node neighbourhoods as proposed here.

# Methods

## Patient group

The patient cohort derived from the one presented by Nissen *et al.*[46] Three cases were removed, two due to the existence of a previous resection, and one due to withdrawal of patient consent. The final patient cohort thus consisted of 91 patients with refractory epilepsy, with heterogeneous seizure aetiology and including both temporal and extra-temporal resections. More details can be seen in Supplementary Table 1. All included patients (i) received a clinical MEG recording as part of their pre-surgical evaluation between 2010 and 2015 at Amsterdam University Medical Center, location VUmc; (ii) subsequently underwent epilepsy surgery at the same centre; and (iii) surgery outcome information was available following the Engel classification[9] either 1 year (88 patients) or at least 6 months (3 patients) after the surgery. Patients who were Engel class 1A and 1B were classified as SF. A waiver of ethical review was obtained from the institutional review board (Medisch Ethische Toetsingscommissie Vrij Universiteit Medical Center) as no rules or procedures were imposed other than routine clinical care.

The patient group was heterogeneous with temporal and extratemporal resections and different aetiology. Surgical outcome was classified according to the Engel classification.[9] Sixty-two patients were deemed SF.

## Individualized brain networks

Individualized brain networks were derived for each patient from 10 to 15 min resting-state MEG recordings, using the Automated Anatomical Labeling (AAL) atlas[64-66] to define a brain parcellation of 90 Regions of Interest (ROIs), with 78 cortical and 12 subcortical ROIs, excluding the cerebellar ROIs.[67] The pre-processing steps, as well as the procedures used to reconstruct the activity of each source, are described in detail in Nissen *et al.*[46] We derived seven brain networks for each patient: a broadband network ($B$, 0.5–48.0 Hz) and six frequency-band specific networks: $\delta$ (0.5–4.0 Hz), $\theta$ (4.0–8.0 Hz), $\alpha_1$ (8.0–10.0 Hz), $\alpha_2$ (10.0–12.0 Hz), $\beta$ (12.0–15.0 Hz) and $\gamma$ (15.0–30.0 Hz), by filtering the source-reconstructed data in the corresponding frequency bands.

Each ROI defined one node in the network, and the coupling strength or link weight between each pair of nodes $w_{ij}$ was estimated with the Phase Lag Index (PLI). The PLI is a functional connectivity metric that measures the asymmetry in the distribution of instantaneous phase differences between two time series.[67] The PLI is insensitive to zero-lag coupling and thus it is robust against volume conduction or field spread.[67] A total of 174 epochs of 4096 samples (3.28 s) were used for each patient to estimate functional coupling.

Raw PLI matrices were thresholded and binarized with a disparity-filter method.[68] The disparity filter extracts the connectivity backbone ($a_{ij} > 0$ if there is a significant connection between $i$ and $j$ and 0 otherwise) of a network by removing connections that are not statistically significant. The disparity filter accounts for node heterogeneity in the edge weight distribution: weak edges are identified on a node-by-node basis, by comparing their strength to that of the remaining node's edges with a given significance threshold $\alpha$, which we set to 0.1. This resulted in sparse networks (with network densities of ∼5%; range: 0.047–0.051, see Supplementary Table 2) with giant components spanning the majority of the nodes (range: 84.49–89.1).

## Local node metrics

We characterized the local structure of the network by three nodal properties. In particular, for each node $i$, we considered its centrality (as given by the betweenness centrality $BC_i$), clustering coefficient $CC_i$ and curvature $C_i$. The 'betweenness centrality' of a node measures its influence over the flow of information on the graph: $BC_i$ indicates the fraction of shortest paths in the network that pass through node $i$. The 'local clustering coefficient' of a node indicates the fraction of triads involving node $i$ that are closed, i.e.

$$CC_i = 2\frac{\text{No. connected triangles including node } i}{k_i(k_i - 1)},$$

$$CC_i = 2\frac{\sum_{1\leq j<l\leq N, j,l\neq i} a_{ij}a_{il}a_{jl}}{k_i(k_i - 1)},$$

where $a_{ij}$ indicates an element $(i, j)$ of the adjacency matrix $A$, which is equal to 1 if $i$ and $j$ are connected by a link or 0 otherwise, $N$ is the number of nodes in the network and $k_i = \sum_j a_{ij}$ is the degree of node $i$. Finally, the local curvature of a network generalizes the concept of curvature of a surface, which intuitively measures how the surface bends in distinct directions[69]:

$$C_i = \sum_{m=1}^{m_{max}} (-1)^{m+1} \frac{Cl_{\Im}}{m},$$

where $Cl_{im}$ is the number of $m$-cliques, to which $i$ belongs, and $m_{\max}$ represents the size (i.e. number of nodes) of the largest clique in the network. Here, we have considered the size of interactions up to three nodes ($m_{\max} = 3$).

## Simplicial complex description

Simplicial complexes represent higher-order networks, which allow for interaction between not only two but also more nodes, described by simplices. A $d$-simplex is formed by a set of $d + 1$ nodes and all their possible connections. For instance, a 0-simplex is simply a node, a 1-simplex is a link and the two corresponding nodes, a 2-simplex is a triangle, a 3-simplex is a tetrahedron and so on. A simplicial complex $K$ is formed by a set of simplices such that (i) if a simplex belongs to $K$, then any simplex formed by a subset of its nodes is also included in $K$, and (ii) given two simplices of $K$, their intersection either also belongs to $K$, or it is a null set.[62] A simplicial complex representation of a network can be built deterministically by defining the *clique complex* of the network. A $k$-clique is a subgraph of the network formed by $k$ all-to-all connected nodes. That is, 1-cliques correspond to nodes, 2-cliques to links, 3-cliques to triangles and so on. Thus, to build a simplicial complex of dimension $d$ from a network, we identify all $d + 1$-cliques.[62,70] This choice for creating simplices from cliques has the advantage of using pairwise signal processing to create a simplicial complex from brain networks.[71] Other strategies to build simplicial complexes beyond pairwise signal processing have been proposed, such as approaches combining information theory and algebraic topology.[63,72-76]

## Extended neighbourhood

The mesoscopic structure of a complex network can be described in terms of extended neighbourhoods or ego networks,[62] as illustrated in Fig. 1. Starting from a given node $i$, we define its $d$-extended neighbourhood $EN_i^d$ as the subgraph induced by the set of nodes at a hopping distance $\delta$ equal or smaller to $d$, $\delta \leq d$ (see Fig. 1B). $EN_i^d$ generalizes the concept of the clustering coefficient, as it allows us to capture the connectivity not only between the first neighbours of a node, but of its general area of influence characterized by the hopping distance parameter $d$.

$EN_i^d$ can be characterized by its size (number of nodes $N_{EN_i^d}$) and connectivity (number of links, $E_{EN_i^d}$). $N_{EN_i^d}$ generalizes the notion of node degree, and indeed the degree of a node equals to $N_{EN_i^{d=1}}$. Similarly, the local clustering coefficient reduces to $CC_i = 2\frac{E_{EN_i^{d=1}}}{N_{EN_i^{d=1}}(N_{EN_i^{d=1}}-1)}$.

Finally, we also characterized the topological organization of the extended neighbourhoods by the notion of 'Betti numbers'. The first Betti number $\beta_0$ measures the number of connected components in a network. Subsequent Betti numbers $\beta_i$ describe the topology of the simplicial complex associated with the network. Generally, the Betti numbers $\beta_i$, $i \geq 1$ are topological invariants derived from the simplicial complex that measure the number of linearly independent $i$-dimensional holes in the simplicial complex. Thus, $\beta_1$ provides the number of 1D cycles that are not boundaries of 2D simplices of the associated simplicial complex, and similarly, $\beta_2$ indicates the number of 2D cycles (i.e. over triangles) that are not boundaries of 3D simplices of the simplicial complex. $\beta_0$ indicates the number of connected components of the local neighbourhood. Thus, large values indicate a hub that connects otherwise disconnected regions of the network.[62] $\beta_1$ indicates the number of cycles forming 1D holes. Therefore, a large value of the ratio $\beta_1/\beta_0$ indicates a sparse neighbourhood. Similarly, larger values of $\beta_2$ indicate the tendency to form planar (i.e. triangular) structures. The Betti numbers are non-linearly influenced by the size and density of the neighbourhood and integrate information on the mesoscopic structure of the network in a non-trivial manner.

## Resection area and node sets

The RA was determined for each patient from the 3-month post-operative MRI. This was co-registered to the pre-operative MRI (used for the MEG co-registration) using FSL FLIRT (version 4.1.6) 12-parameter affine transformation. The RA was then visually identified and assigned to the corresponding AAL ROIs, namely those for which at least 50% of the centroid had been removed during surgery.

Based on the RA, we identified four sets of nodes: $RA$, or resected nodes, are the nodes that belong to the RA. $\overline{RA}$, or non-resected nodes, are the nodes that do not belong to $RA$. We further considered two subsets of $\overline{RA}$. This partition was based on the connectivity of the RA and was thus different for each frequency band: $N$, or neighbours, are the nodes that are connected to $RA$ nodes and that do not themselves belong to the RA. $O$, or other nodes, are the remaining nodes in the network, that is, nodes that do not belong to the RA and are not connected to any $RA$ nodes. In Supplementary Fig. 1, we report the distribution of the sizes of each node set, for each frequency band, over the patient cohort.

## Statistical analyses

We first performed an individualized, node-based analysis, by which we tested whether the hub status of the different node sets differed significantly for each patient and metric $X$. We considered two types of comparisons: (i) two-node-set setting, where we tested whether $X(RA) > X(\overline{RA})$, and (ii) three-node-set setting, where we tested whether $X(RA) > X(N)$, $X(RA) > X(O)$ and $X(N) > X(O)$. We quantified whether the hubness distributions were significantly different via bootstrapping analyses to determine the $z$-score and *P*-value of the difference. In particular, the statistical significance of the difference between the two data sets was assessed by performing $10^4$ bootstrapping replicas of each data set, which were drawn randomly, independently and with replacement from each data set. For each pair of samples, we calculated the difference between their means. From the resulting distribution of mean differences, we computed the mean and standard deviation, deriving a *z*-score and corresponding two-tailed *P*-value. The sign of the difference indicated whether it was in the direction of the hypothesis or against it. The *z*-score was computed as the mean of the differences of the bootstrapped samples divided by the standard deviation of these differences. The two-tailed *P*-value associated with the *z*-score was determined using the cumulative distribution function of a standard normal distribution. Considering the large number of comparisons performed, we applied the Bonferroni correction to account for multiple testing and control the false discovery rate (FDR). Specifically, the Bonferroni correction was applied by dividing the original significance level ($\alpha = 0.05$) by the number of comparisons made for each patient ($n = 56$, seven frequency bands times eight generalized centrality metrics), resulting in the new significance level $\alpha' = \frac{0.05}{56} \approx 8.9 \cdot 10^{-4}$.

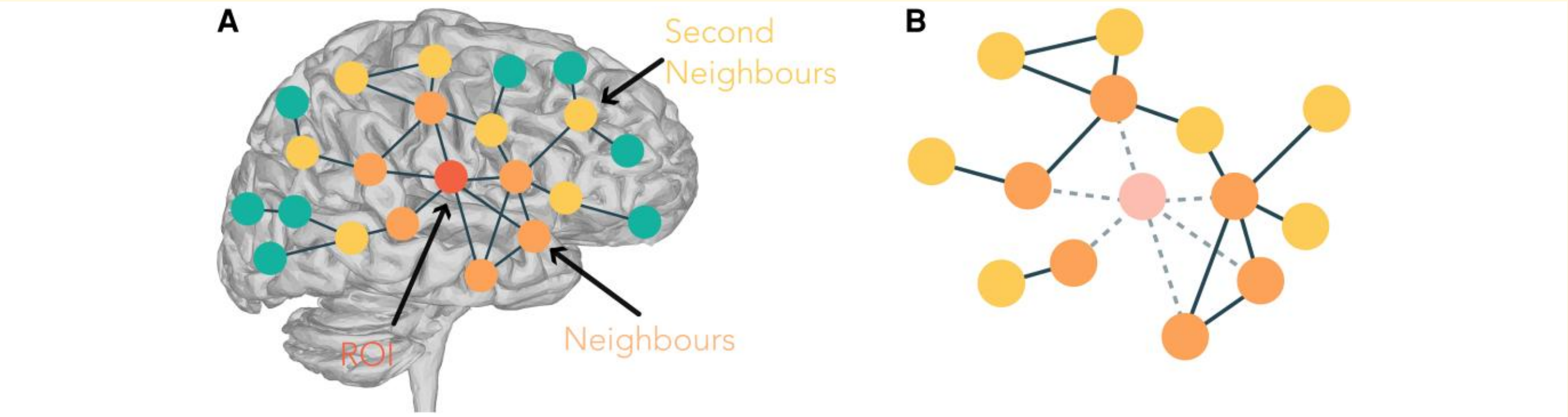

**Figure 1 Schematic description of extended neighbourhoods.** (**A**) Illustrative representation of the extraction of a node's neighbourhood for $d = 2$. The nodes are colour-coded to show the ROI (central node, shown in red), its first (orange) and second (yellow) neighbours and the remaining network nodes (green). (**B**) Extended neighbourhood of the node. The central node is not included in its neighbourhood, therefore it is shown here with low opacity (light pink node) and its edges are removed (dashed lines). The topological organization of the neighbourhood can be observed. In this case, e.g. two different connected components emerge, as well as two closed triangles.

To determine whether the results held at the group level, for each patient, we estimated the average generalized centrality of each of the four node sets, for each frequency band and network metric. Then, we performed pairwise statistical comparisons of the node sets at the population level. Therefore, each test comprised $n = 91$ data points, one for each patient. The difference was measured via the *z*-score and *P*-value of a paired bootstrapping analysis, performed as described above. The *P*-values were then Bonferroni corrected to account for multiple comparisons ($n = 56$).

We subsequently utilized the results of the node-based analyses to perform a receiver operating characteristic (ROC) curve classification of the patients [SF or non-seizure free (NSF)]. The result of each node-based test was quantified in the variable $r^i_{n1,n2}(X)$ for each patient *i*, hubness metric *X* and node sets $n_1$ and $n_2$. $r^i_{n1,n2}(X) = 1, \ -1$ or 0 indicating whether the node sets were significantly different in the direction of the hypothesis, contrary to it, or not significantly different, respectively. We then summed over hubness metrics to define a ‘distinguishability’ score $D^i_{n1,n2}$ for each patient and node-based comparison.[43] To sum up the results of the three-node-set analysis, we defined a combined distinguishability score, $D^i_{comb}$, by summing over the corresponding three pairwise comparisons. The distinguishability score according to each test was then used to classify the patients with a ROC curve analysis, and the goodness of the classification was measured with the AUC. We performed a bootstrap analysis to derive confidence intervals for the AUC values. For each frequency band and type of comparison, we created $n = 10^4$ bootstrap samples by resampling the original data with replacement and performed the ROC curve analysis on each sample. This resulted in a distribution of $10^4$ bootstrap AUC estimations, from which we derived 90% confidence intervals. To assess whether the AUCs of different tests (i.e. type of comparison and frequency band) differed significantly, we compared the measured AUC of one test against the bootstrap distribution of the other, for all test pairs. Let *Bi* and *Bj* denote the sets of bootstrap replicates for tests *i* and *j*, $i \neq j$, with respective AUCs denoted as AUC*i* and AUC*j*. We computed the fraction of bootstrap samples *Bi* smaller than AUC*j*, and vice-versa, the fraction of bootstrap samples *Bj* smaller than AUC*i*. The average yielded a two-sided *P*-value for the null hypothesis that there was no significant difference between AUC*i* and AUC*j*. The *P*-values were then FDR corrected for multiple comparisons with the Benjamini–Hochberg procedure, and the significance threshold was set at 0.05.

Finally, to enable a more direct comparison with the previous study by Englot *et al.*,[43] we also estimated the distinguishability score as originally proposed, by calculating the AUC of the node ROC classification (instead of using $r^i_{n1,n2}(X)$), for each metric *X* and pair of node sets. Patient classification based on this score was then performed similarly to the previous analysis. For this test, we also considered the node strength (the sum of its non-zero weights after thresholding) as a metric to allow for a more direct comparison with Ramaraju *et al.*[48]

# Results

We considered a cohort of 91 epilepsy-surgery patients derived from a previous study.[46] The cohort included patients with both temporal and extra-temporal resections (see Supplementary Table 1 for details, including patient demographics). For each patient, we derived their resting-state functional brain connectivity from MEG recordings (see Methods for details), considering the AAL (Automated Anatomical Labelling) atlas ($n = 90$ ROIs) and the PLI connectivity metric.

## Extended neighbourhoods

To characterize regional brain organization, we have considered the notion of the extended neighbourhood *EN* of a node.[62] Extended neighbourhoods, also called ego-centred networks, define the area of influence of a node. Mathematically, the extended neighbourhood of node *i*, $EN^d_i$, is defined as the subgraph formed by nodes at distance $\delta$, $0 < \delta \leq d$, of node *i* (which, crucially, excludes node *i*, see Methods for a detailed definition), as depicted in Fig. 1. By changing the radius *d* of the extended neighbourhood, we can access different scales of network organization, going from the local to the global perspective. To quantify the structure of $EN^d_i$, and thus regional network organization, we have considered five topological measures: the size or number of nodes $N^d_i$, the number of edges $E^d_i$ and the first three Betti numbers, quantifying the number of connected components $\beta^d_{0,i}$, the number of loops $\beta^d_{1,i}$ (not accounting for triads, which are always considered to be filled, see Methods) and the number of cavities or 2D loops, $\beta^d_{2,i}$. These metrics quantify the

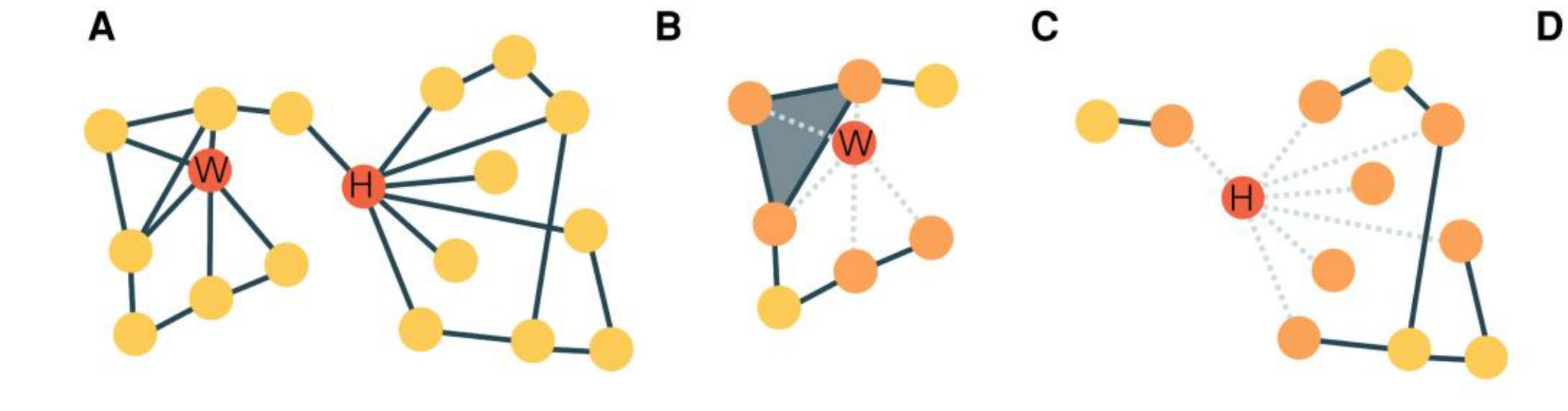

D

| Metric | W | H |
|---|---|---|
| $E$ | 7 | 11 |
| $N$ | 7 | 7 |
| $\beta_0$ | 1 | 4 |
| $\beta_1$ | 1 | 0 |
| $\beta_2$ | 0 | 0 |
| $BC$ | 0.2 | 0.7 |
| $c$ | 0.4 | 0 |
| $\mathcal{C}$ | 1.08 | -2.5 |

**Figure 2 Illustration of the properties of simplicial complexes and extended neighbourhoods.** (**A**) Schematic network where we highlight two nodes: a regional hub (node *H*) with a high degree (seven neighbours) and high Betweenness Centrality *BC* since it brokers two communities, and a local hub (node *W*) with a high degree (five neighbours) but low *BC*. (**B**) and (**C**) illustrate extended neighbourhoods of *W* and *H*, respectively. The different topology of the extended neighbourhoods $EN_W$ and $EN_H$ is encoded by the regional and local metrics, as shown in (**D**). Whenever there is a closed clique in the original network, simplices are built in the extended neighbourhood. For instance, in **B**, the grey triangle represents a 2D simplex built according to this rule.

topology of the extended neighbourhood of each node. The number of nodes and edges indicates the regional connectivity of the node and can be interpreted as centrality metrics. Similarly, a high value of the first Betti number indicates that node *i* acts as a broker between different otherwise disconnected components of its neighbourhood.[62] In Fig. 2, we provide an illustration of these different metrics. A more detailed description of each metric can be found in the Methods and in the Supplementary Material.

As a benchmark, we have also considered three node-based metrics, namely the betweenness centrality $BC_i$, the local clustering coefficient $c_i$ and the local curvature $C_i$. The betweenness centrality is a standard measure to quantify node-centrality and define *hubness*.[21,29,45] It quantifies the extent to which a node lies on the shortest paths between other nodes, thus capturing its role in controlling information flow in the network.[77] Node curvature measures how paths in the simplicial complex diverge or converge around a node, capturing the local geometric properties of the space. Specifically, in a simplicial complex, curvature reflects how higher-dimensional simplices (such as triangles or tetrahedra) connect around a node, influencing the shape and flow of the network structure. It is associated with network robustness and also identifies brain hubs, with large negative values being indicative of hub status.[78] The clustering coefficient captures the connectedness of a node's neighbours and has previously been associated with epilepsy-surgery outcomes.[20]

## Topological characterization of the epileptogenic zone

Following the hypothesis that the EZ is either a hub or connected to a hub, we hypothesized that RA nodes and their neighbourhoods will be more central than other nodes in the network. To test this hypothesis, we considered an existing database comprising 91 patients who underwent epilepsy surgery at Amsterdam UMC, location VUmc. This database had been studied with a combination of network metrics and machine learning previously.[46] The brain organization for each patient was encoded in a functional brain network comprised of 90 ROIs (according to the AAL atlas[64]), derived from resting-state MEG, and thresholded to keep only the strongest links (see Methods for details). MEG networks were derived in different frequency bands, which account for different aspects of brain function. For simplicity, we have considered here first the broadband (0.5–48.0 Hz) but refer back to a multi-frequency analysis in later sections. The RA of each patient was derived from post-operative MRI and was encoded in terms of AAL nodes. We note that this hypothesis is better-suited for SF patients, for whom the EZ is known to be included in the RA, than for NSF patients, for whom we know that at least part of the EZ remained unresected. Therefore, besides whole-cohort analyses, we analysed the SF and NSF subgroups independently.

Each node in the network was described by means of the 8 metrics defined in the previous section, with high values of these metrics associated with higher generalized centrality, except for the curvature, where the direction is the opposite as discussed above. Initially, two sets of nodes were defined for each patient and network: resected nodes *RA* and non-resected nodes $\overline{RA}$. We analysed whether these nodes differed at the individual level in any of the 8 metrics considered (bootstrapping and a Bonferroni correction were used to establish statistical significance, see details in the Methods). Details of this analysis for an exemplary case are shown in Supplementary Fig. 2. We found that, at the individual level, *RA* nodes were significantly more central according to all neighbourhood metrics, except $\beta_0$, for 15–30% of patients (respectively for 23, 27, 20 and 14 cases for $N$, $E$, $\beta_1$ and $\beta_2$). Traditional node-based metrics were less efficient at detecting differences between the node groups: according to these metrics, *RA* nodes were significantly more central than $\overline{RA}$ nodes only for a handful of patients (respectively 6, 3 and 6 for $c$, $C$ and $BC$). We note that, for a few patients, the opposite result was found and *RA* nodes were significantly less central than $\overline{RA}$ nodes, both with the node- and neighbourhood-based metrics (respectively 5, 5, 7 and 3 cases for $N$, $E$, $\beta_1$ and $\beta_2$; and 3 and 1 cases for $c$ and $C$; whereas no case was found for $BC$). These results are summarized in Fig. 3, Panel A, whereas numerical results can be found in Supplementary Tables 3–5. On average (across network metrics and frequency bands), 5.02% of patients had a significant difference in the direction of the hypothesis (positive), and 2.32% in the direction against it (negative). We found that these results were replicated both in the SF and the NSF subgroups (see Supplementary Fig. 3 and Supplementary Tables 3 and 4) with similar dependencies on the type of node-test, frequency band and generalized centrality metric.

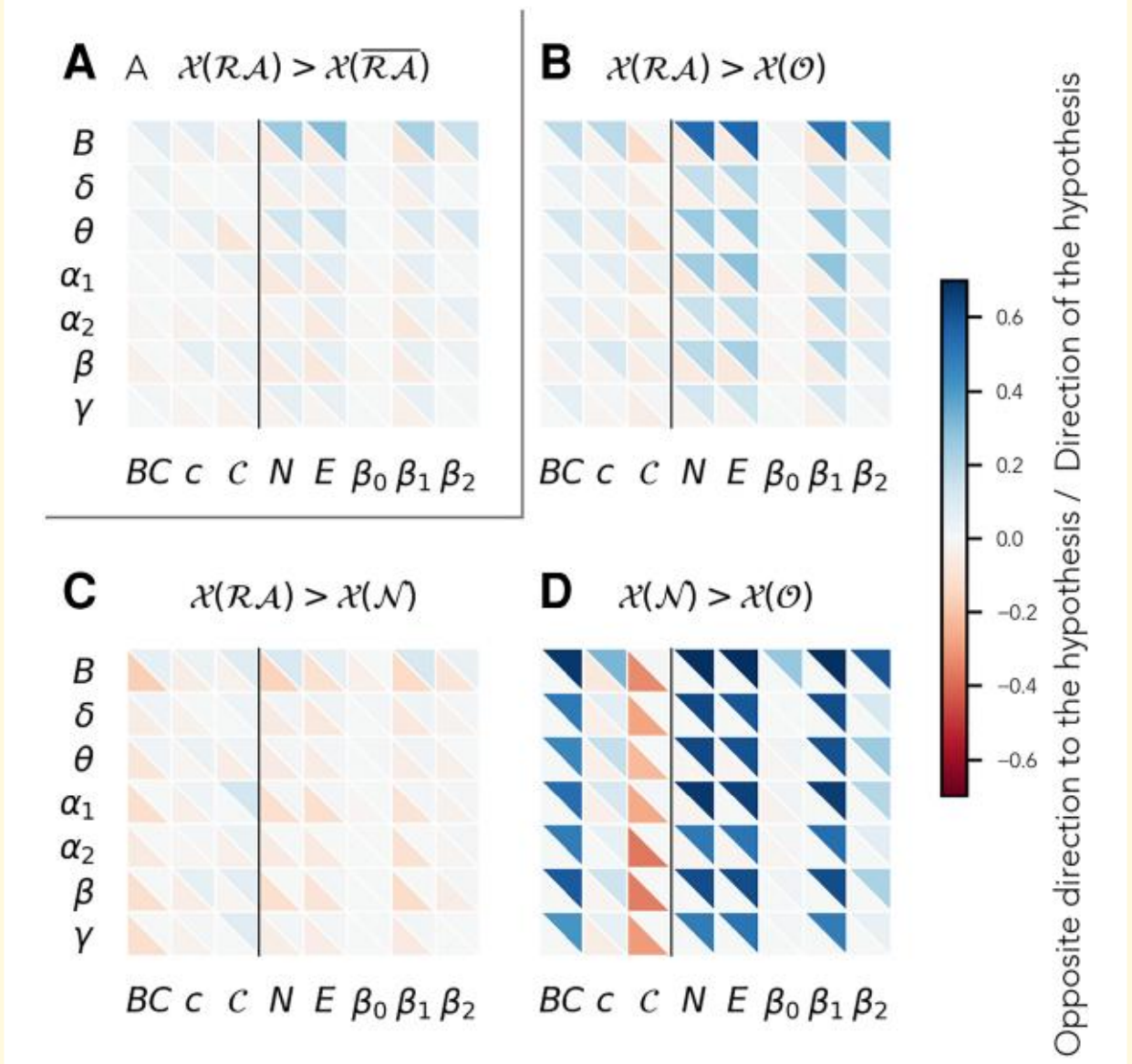

**Figure 3 Results of the patient-specific node-type comparison.** Patient-specific comparison of different node groups for the two-group (**A**) and the three-group (**B**)–(**D**) set-ups. For each panel, the hypothesis of the relation in centrality between the two groups is shown in the panel title. The fraction of patients for whom there was a significant difference in the direction (opposite direction) of the hypothesis is shown by the blue (red) triangles in the upper-right (bottom-left) corner of each cell, respectively, for each frequency band (rows) and metric (columns), colour-coded as indicated by the colorbar. The corresponding numerical values are shown in Supplementary Tables 3 and 4. The vertical black line separates node-based (left) from neighbourhood-based (right) metrics. $X$ ($S$) stands for the generalized centrality metric $X$ measured on the nodes in set $S$. Significance (threshold = 0.05) was established with a bootstrapping analysis (see Methods, $n$replicas = 104). The number of samples in each test was different for each patient since it was determined by the size of each node set. The $P$-values were Bonferroni corrected for multiple comparisons ($n = 56$).

Our results agree with previous studies, according to which the RA is not always a network hub, but it is often strongly connected to a pathological hub.[21,29,46] Consequently, the $\overline{RA}$ node set may include both nodes that are less and more central than $RA$ nodes. To take into account this effect, we split the $\overline{RA}$ set into two: nodes that were neighbours of the $RA$ (*neighbours*, $N$ set) and nodes that were not (*other*, $O$ set). According to our initial hypothesis, within this division of the node sets, we expected that both $RA$ and $N$ nodes were more central than $O$ nodes, and that $RA$ and $N$ nodes were similarly highly central. As expected, we found that RA nodes were significantly more central than O nodes for 13.80% of patients on average (over frequency bands and network metrics), whereas the opposite was true only for 2.10% of patients. Similarly, $N$ nodes were significantly more central than O nodes for 37.94% of cases, whereas the opposite was true only for 0.60% (Fig. 3, Panels B and D). As before, neighbourhood-based metrics were able to capture this difference more consistently across patients than node-based metrics. Regarding the relative hub status of the RA and its neighbourhood, we only found significant differences between $RA$ and $N$ nodes for a small fraction of the patients (1.66% in the positive direction and 5.75% in the negative direction, see Fig. 3, Panel C). These went in both directions, with a tendency towards a higher centrality of $N$ nodes at the group level, as we discuss below. For instance, for the metric that picked up the most differences in the broadband, $\beta_1$, $RA$ nodes were more central than $N$ nodes for 10 cases, but the opposite was true for 12 cases. These findings indicate heterogeneity in the patient population regarding the relative hub status of the RA and its neighbours. For most cases, these two sets could not be distinguished based on centrality metrics (either node- or neighbourhood-based), indicating a similar highly-central status (note that the remaining nodes were found to be less central).

## Group-level analyses

To gain a population-level perspective of the relative hub status of the RA, we repeated the previous analyses at the group level. To do so, we measured the average centrality of the nodes in each node set, for each patient and frequency band. We found that, when all patients were pooled together, the differences between node sets became more subtle, likely due to patient-specific variability, as shown in Fig. 4 (numerical data are reported in Supplementary Tables 6 and 7). Overall, we found in the two node-group analysis that the $RA$ and $\overline{RA}$ node sets could not be significantly distinguished at the group level, for most metrics and frequency bands, with the most notable exception of the broadband. The three node-group analysis recovered for the most part the findings of the individual-level analyses, i.e. $O$ nodes were the least central, and $N$ were somewhat more central than $RA$. At the group level, the betweenness centrality became the most robust metric across frequency bands, and the broadband network was the network for which differences between node groups were more prevalent across metrics. Notably, three of the metrics, the local clustering $c$, $\beta_0$ and $\beta_2$, performed poorly for the remaining frequency bands.

We repeated this analysis on the SF and NSF subgroups separately (see Supplementary Fig. 4 and Supplementary Tables 8–11). The main findings remain the same in both subgroups, with some loss of significance, which is expected due to the reduced group sizes, particularly for the NSF group. Notably, most changes appear in the $X(RA) > X(N)$ and $X(RA) > X(O)$ comparisons for the NSF group. In the first case, no significant differences are found in the slow frequency bands and the broadband. In the latter, most significant differences vanish, particularly for the fast frequency bands. This result suggests that weaker differences might be found for the NSF group under the three-node-set partition, at least at the group level. We will explore this result in more detail in the next section.

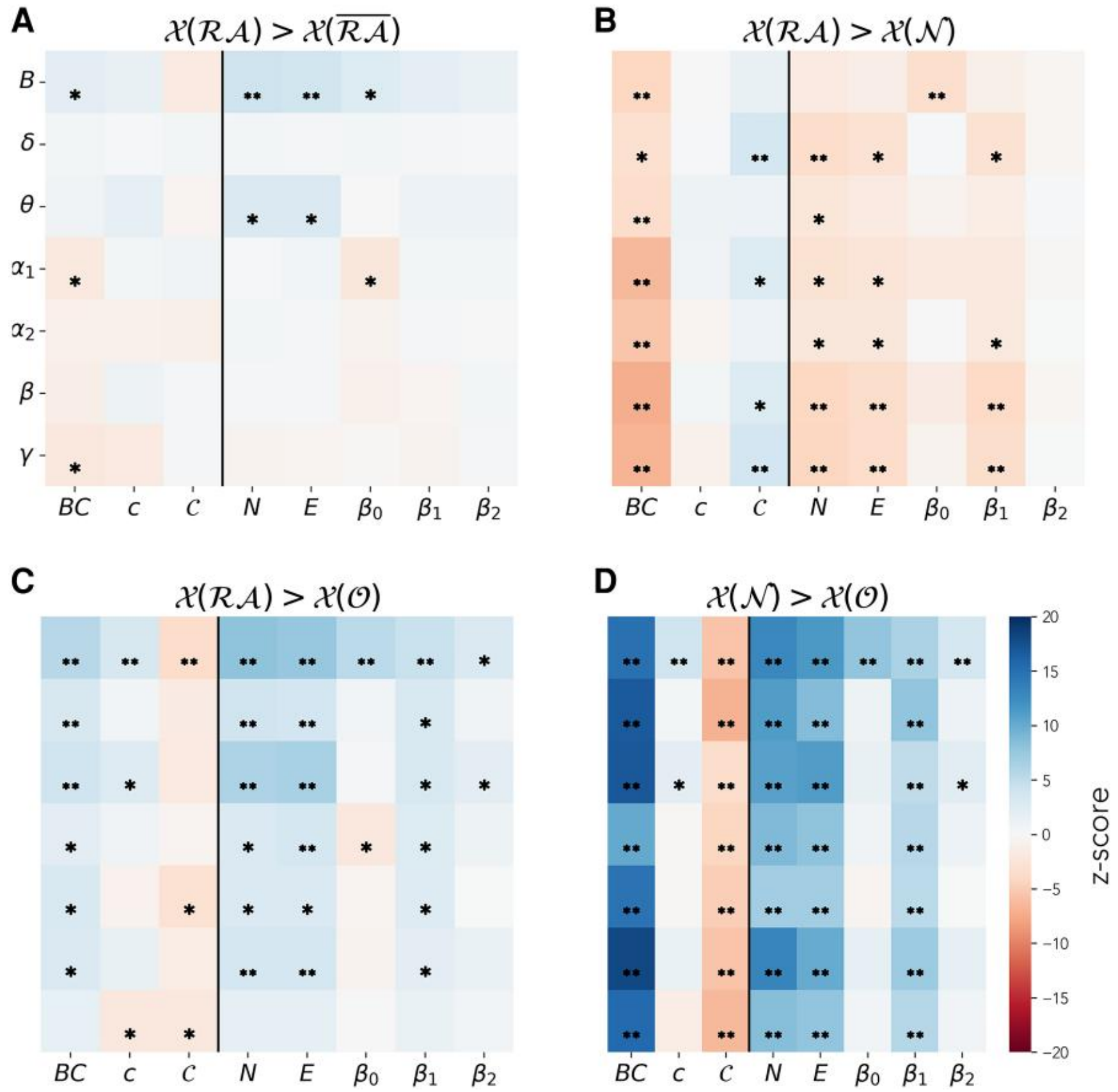

**Figure 4 Results of the group-level comparison of node types.** Group-level comparison between node sets, for each considered frequency band (*y*-axis) and network metric (*x*-axis). From left to right, the panels indicate the difference between the node sets: (i) *RA* versus $\overline{RA}$ (**A**), (ii) *RA* versus *N* (**B**), (iii) *RA* versus *O* (**C**), (iv) *N* versus *O* (**D**). *X* (*S*) stands for the generalized centrality metric *X* measured on the nodes in set *S*. The colour code indicates the z-score of the difference between the average values of each node set, computed by bootstrapping the data (sampling size of $10^4$). Single asterisks indicate significant differences ($P < 0.05$, according to the bootstrapping analysis) that did not survive the Bonferroni correction ($n = 56$, see Methods), and double asterisks the ones that did ($P < 8.9 \cdot 10^{-4}$). The number of paired samples in each test was equal to the number of patients, $N = 91$. The corresponding numerical values are shown in Supplementary Tables 6 and 7.

## Topological signatures of the resection area and surgical outcome

To investigate whether the hub status of the *RA* was associated with surgical outcome in this dataset, we assigned each patient a 'distinguishability' score $D_{RA,\overline{RA}}$ to quantify the distinguishability between the *RA* and $\overline{RA}$ node sets.[48]

For each patient, $D_{RA,\overline{RA}}$ measures the number of tests (over the eight network metrics considered) for which the hypothesis of the hub status of the *RA* is significantly fulfilled (see Methods for details on data and how the statistical analyses have been performed, see also Supplementary Fig. 5). Following our previous findings that a three-node-group division is more informative at the node level, we also assigned distinguishability scores to the pairwise comparisons between the three node sets *RA*, *N* and *O*, namely $D_{RA,N}$, $D_{RA,O}$ and $D_{N,O}$. Next, to summarize the results of the three-node-group analysis into one score, we defined a combined score of the three-node-group analysis $D_{\text{comb}}$ by summing over the corresponding three pairwise comparisons (see Methods for details). The derivation of these metrics is illustrated in Supplementary Fig. 3. We used each distinguishability score to classify the patients between the SF and NSF groups, as shown in Fig. 5 (a statistical comparison between the two groups was also performed, see Supplementary Fig. 6, but no statistical differences between the two groups survived after FDR correction). We found that patient classification was fair at best for any of the frequency bands or node-group montages (Panel C, see also Supplementary Table 12 for the numerical values and 90% confidence intervals, and Supplementary Fig. 7 for the exemplary ROC curves for the broadband). There were differences in AUC values depending on the frequency band and type of comparison considered. As shown in Supplementary Fig. 8 (and as it can be inferred also from the data in Fig. 5C), for a given type of node comparison, main differences appear between the slow and fast frequency bands, with slow frequency bands (except for the δ band) showing in general higher AUC values (>0.5), and fast frequency bands showing lower AUC values (smaller than 0.5). These results suggest that the hub distribution of SF and NSF cases changes differentially with the frequency band. The best results were found for the broadband when considering the combined information of the three-node-group analyses, which resulted in an AUC = 0.68.

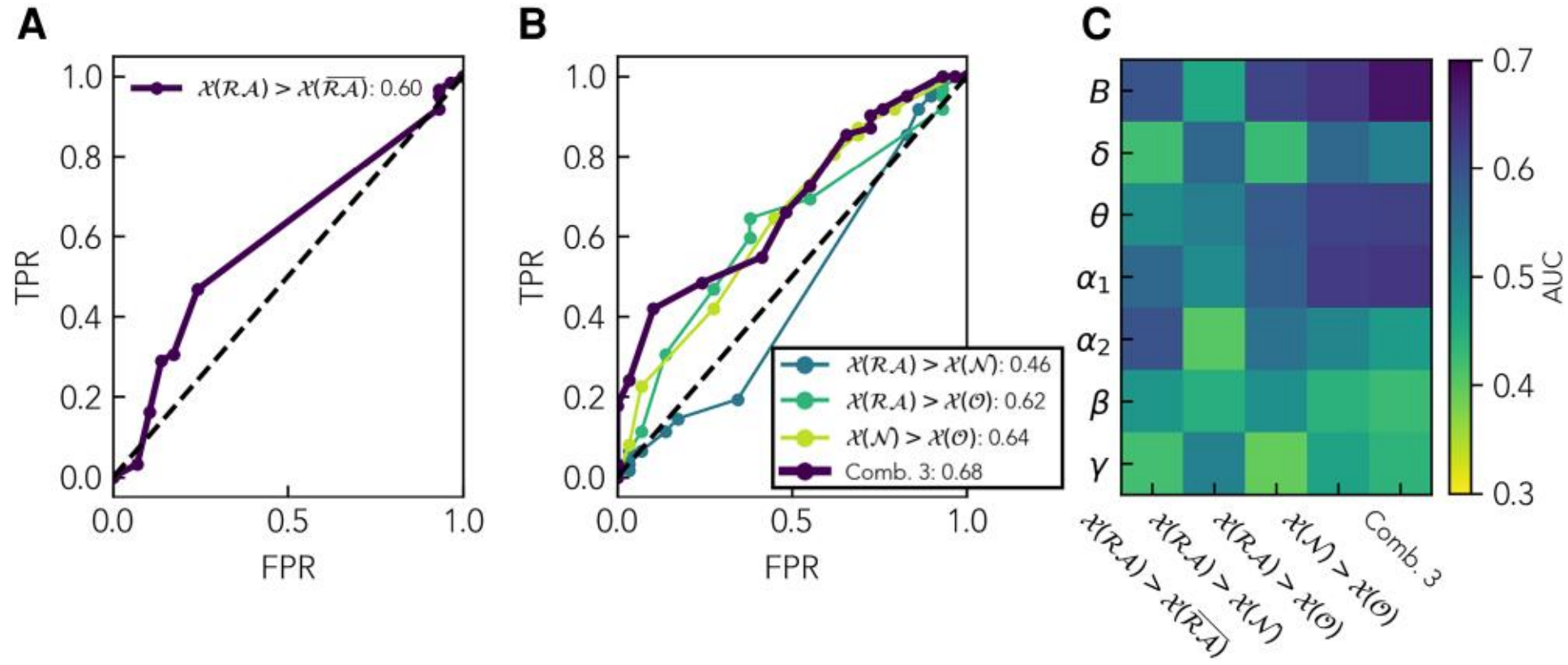

**Figure 5 Results of the patient classification analyses.** Classification of SF and NSF patients based on the two-node-group (**A**) and three-node-group (**B**) distinguishability scores. $X$ ($S$) stands for the generalized centrality metric $X$ measured on the nodes in set $S$. **A** and **B** show the ROC curves corresponding to the broadband, showing the True Positive Rate (TPR) as a function of the False Positive Rate (FPR) for each classification threshold; the remaining bands are shown in Supplementary Fig. 7. The resulting AUC is indicated by the figure legends. **C** shows the resulting AUC for all frequency bands for this same analysis. In this representation, the SF group is assigned to be the positive class. The colour scale is centred around AUC = 0.5, which indicates a lack of association. Blue colours stand for an association in the direction of the hypothesis (AUC > 0.5, i.e. the SF group presents a higher distinguishability score), whereas yellow colours stand for the opposite (AUC < 0.5, the NSF group presents a higher distinguishability score).

We analysed whether there were differences in the patient-classification analysis between patients with temporal and extra-temporal resections (see Supplementary Fig. 9 for details). We found that patients in the extra-temporal subgroup could be classified significantly better overall, particularly for some specific tests. No clear dependency of the difference in the classification on the frequency band or type of node comparison was observed. Similarly, there were no significant differences at the individual test level (after FDR correcting for multiple comparisons).

Finally, to better contextualize and validate our findings, we considered an alternative definition of the node distinguishability, $D'$, as introduced by Ramaraju *et al.*[48] In this case $D'$ is simply the AUC resulting from the classification of $RA$ and $\overline{RA}$ nodes (see Supplementary Fig. 6). In their original study, Ramaraju *et al.*[48] found that they could classify the patients according to surgical outcome with an AUC of 0.76 using $D'$ based on the degree-centrality as a metric. For our dataset, however, we found a maximum AUC of only 0.65 when using $D'$ based on the degree-centrality (see Supplementary Table 13). When applying $D'$ to the 8 metrics considered in the main part of this study, we found AUC values ranging from 0.68 (for the neighbourhood metric $N$ in the $\alpha_1$ band) and 0.36 (neighbourhood metric $\beta_1$, $\alpha_1$ band) for the two-node-set analysis, with similar results also for the three-node-set partition (see Supplementary Figs 10 and 11).

# Discussion

In our study involving 91 patients who underwent epilepsy surgery, we investigated the hub status of the RA and its region of influence to shed new light on the presence of pathological hubs in the brains of epilepsy-surgery patients and their role in the outcome of epilepsy surgery. We proposed a novel methodology based on node-neighbourhoods and topological data analysis to quantify node centrality at a mesoscopic level. As a validation of our novel approach, we compared our findings against established node-based metrics such as the betweenness centrality and clustering coefficient. Moreover, by leveraging the same database previously analysed by Nissen *et al.*[46] with traditional methodologies, we enabled a direct comparison between the two studies.

In our study, we found that (i) the neighbours of the RA play an important role in brain-network organization in epilepsy and are significantly different from the remaining nodes in the networks (thus a three-group partition of the brain regions, where RA neighbours are separated from the remaining brain network, is more representative than a two-group partition); (ii) the RA and its neighbours are more central than the remaining brain regions, which holds true at the group level and also individually for most patients, and both for patients with good and bad outcome; (iii) the RA and its neighbours are similarly highly central, with only some differences at the individual level (for 10–20% of patients) that go in both directions, whereas at the group level the neighbours are weakly but significantly more central; and (iv) the difference in hub status between either the RA or its neighbours and the remaining network nodes, but not between them, is weakly associated with surgical outcome (AUC = 0.62, 0.64 and 0.46, respectively). A main consequence of our findings is that a three-node-group partition of the brain regions, as we have introduced here, such that the RA-neighbouring regions are separated from the remaining brain regions and considered specifically, is more representative than a two-node-group partition, in particular yielding better node-classification results. These findings support the hypothesis of the emergence of pathological hubs in refractory epilepsy that do not necessarily overlap with the RA, a finding that was valid for patients with good and bad outcomes. These results further highlight the need for individualized studies that take into account patient-specific brain connectivity.

## Hub status of the resection area

In this study, we considered the emergence of pathological hubs in epilepsy and their overlap with the RA.[29,45] The RA has been associated with brain hubs both in functional and structural studies[15,22,43-48,57-59,79,80] (see also Stam[29] and Xu *et al.*[81] for recent reviews), and their overlap has been related to surgical outcome, with several MEG studies finding that hub removal was associated with good post-surgical outcomes.[45,47,48,80] In particular, Nissen *et al.*[45] found that the brain network hubs [defined via the betweenness centrality on a minimum-spanning-tree (MST), description] were localized within the resection cavity in 8 out of 14 SF patients and none (out of 8) NSF patients (73% accuracy). Similarly, Fujiwara *et al.*[80] found that removal of the most central hubs (defined via the eigenvector centrality on weighted PLI networks) had predictive value in a study with 31 patients (17 SF). Considering a simple correlation metric as the basis for connectivity, Ramaraju *et al.*[48] found, in a study with 31 patients (12 SF), that SF patients had significantly more hubs surgically removed. Finally, Corona *et al.*[47] also found higher functional connectivity (defined via both the amplitude-envelope coupling and phase-locking-value on the MST description) inside than outside the RA for SF patients, and a few differences between the two for NSF patients in a study with 37 (22 SF) patients involving both children and young adults with refractory epilepsy. The functional connectivity measures predicted weakly the EZ location and surgical outcome (sensitivity and specificity > 0.55 with leave-one-out cross-validation).

However, the relationship between hub-removal and surgical outcome could not be validated in our previous study[46] (94 patients, 64 SF), which used the same patient cohort as we have considered here. Nissen and colleagues defined the hub status on the basis of the MST betweenness centrality, and only a weak association with the RA was found (60.34% accuracy with a random forest classifier) and none with surgical outcome (49.03% accuracy). In line with the suggestion that the relationship between the RA and the brain hubs is not straightforward, several studies have pointed towards the functional isolation of the EZ, both in invasive EEG[38] and MEG.[39] In particular, Aydin *et al.*[39] found that SF patients presented a more isolated RA (relative to the contralateral hemisphere) than NSF patients in a study with 12 patients (7 SF) based on amplitude-envelope-correlation networks. Johnson *et al.*[38] found that the SOZ and the early propagation zone presented increased inwards and decreased outwards functional connectivity in an invasive EEG study involving 81 drug-resistant epilepsy patients undergoing pre-surgical evaluation. Interestingly, they found that the largest difference between SF and NSF patients appeared in the propagation zone: the connectivity profile of the propagation zone was intermediate to that of the SOZ and the remaining networks for SF patients, whereas for NSF patients it consistently and closely resembled that of the remaining network. It is worth noting that this result may just reflect a difference in invasive EEG sampling between SF and NSF patients, such that, e.g. the true propagation zone of NSF patients may have been undersampled.[38]

The existence of pathological hubs can reconcile these findings: a pathological hub that may or may not coincide with the SOZ may be present, facilitating seizure propagation. Then, removal of either the SOZ, the pathological hub, or even the connection between them could lead to seizure freedom.[24,45] In a previous modelling study, for instance, we found that the link-based resections that led to the best post-surgical outcome in the model were those linking the RA to the network hubs.[21] Our findings in the current study support this hypothesis, as we have found that the relative hub status of the RA varies largely within the patient cohort, and that whether it is more or less central than its neighbours does not determine outcome. Therefore, removal of a hub region was not necessary in this study to achieve seizure freedom. Of note, in this study, we have considered the RA as a proxy for the EZ, as is commonly done in epilepsy-surgery studies.[23,46,48] However, this adds a level of inaccuracy: for NSF cases, it is known to be inaccurate, but even for SF cases, it might have been larger than needed.[21,22] This can lead to inaccuracies in the definition of the RA, and as a consequence in the neighbourhood regions. In contrast, the differences between either the RA or its neighbours with the remaining brain regions proved to be a stronger indicator of surgical outcome (albeit still weak, with AUC = 0.62 and 0.64, respectively). The proposed three-node-set partition may thus provide new insight into the effect of a particular resection, which may be missed with the standard two-node-set partition approach. This is in agreement with the methodology and findings by Johnson *et al.*,[38] but here we propose a methodology based only on resting-state MEG brain connectivity, without the need for invasive or ictal recordings, as the notion of the propagation zone is substituted by that of the neighbours of the RA. Given that the type of resection (temporal versus extra-temporal) plays a major role not only in surgical outcome but also in the expected centrality of the RA, we analysed whether there were differences between the temporal and extra-temporal subgroups regarding the patient-classification analyses. We found that, overall, the extra-temporal subgroup yielded larger AUC values than the temporal subgroup. This finding suggests that centrality measures might provide more information on the expected surgical outcome for patients with extra-temporal resections. This is the subgroup with the worse prognosis (only 50% were SF in this data set, relative to 77.05% of patients with temporal resections, see Supplementary Table 1), and which thus could benefit the most from innovative analyses. This can be a promising line of research for further studies that focus on the population of patients with extra-temporal resections.

Finally, we note other possible interpretations of our finding that pathological hubs are associated with the RA in epilepsy surgery. Firstly, the emergence of pathological hubs in the functional brain networks of epilepsy patients might be driven by the presence of interictal epileptogenic activity, such as spikes or high-frequency oscillations, which can drive connectivity metrics. The measure considered here as an indicator of functional connectivity, the PLI, should minimize such effects by considering only phase coherence and neglecting zero-lag coupling. Secondly, the association between the RA and network hubs might be driven by actual network hubs—not pathological ones. This poses a general problem in epilepsy-surgery planning, as resections with the best outcome—by targeting the brain hubs—may also have important side-effects.

## Centrality metrics and node neighbourhoods

In this study, we proposed the use of regional centrality metrics to better account for the effect of a given resection, following previous theoretical works.[62,82,83] Most previous clinical studies have considered traditional centrality metrics that do not take the local network-neighbourhood into account, of which the degree,[45-48] betweenness centrality,[45,46] and eigenvector centrality[21,84,85] are predominant. Here we found that neighbourhood-based metrics, with the exception of $\beta_0$ (which equaled 1 in most cases for the considered parameters, as a consequence of the high level of recurrent connectivity in the networks), were able to more consistently pick up differences between $RA$ and $\overline{RA}$ nodes at the individual level across all frequency bands, and in particular for the broadband, than nodal measures such as the betweenness centrality or the clustering coefficient (Fig. 3). These findings indicate that the neighbourhood of the RA is significantly different from the neighbourhood of other nodes in the brain network, in particular denoting a higher (generalized) centrality. In contrast, at the group level (Fig. 4), the metric that revealed the strongest difference between $RA$ and $\overline{RA}$ nodes was the betweenness centrality, which is also the metric most often considered in the literature. We note this as an interesting avenue for future research: at the theoretical level to understand whether different centrality metrics might be more or less sensitive to individual variations, and at the clinical level to validate the generalizability of these findings. Notably, whereas the betweenness centrality requires global information, extended neighbourhood metrics can be computed with only regional information, and are thus more efficient to compute for large systems.

In the case of the three-node-set partition, at the group level, the differences between local and regional centrality metrics were larger (Fig. 4 and Supplementary Table 6). This may be caused by the neighbourhood-based partition of the node sets, such that the RA neighbourhood is considered explicitly, even for the node-based metrics. At the individual level, the neighbourhood-based metrics were also slightly more sensitive to differences between both the RA ($RA$ set) and its neighbours ($N$ set), with the remaining network nodes ($O$ set). Differences between the $RA$ and $N$ node sets were sparse as discussed above, and generally all metrics performed similarly except for the clustering coefficient $c$, and the first and third Betti numbers, $\beta_0$ and $\beta_2$, with very low sensitivity. In particular, $\beta_0$ and $\beta_2$ showed little variation across nodes for the parameters considered. At the group level, however, the betweenness centrality and curvature found the strongest and most consistent differences between node sets. Further studies, considering e.g. larger networks or different connectivity thresholds, could validate the generalizability of these findings.

To better contextualize our study, we also considered the node strength (or weighted degree) as a centrality metric, following Ramaraju *et al.*[48] In their original study the authors found that this metric could classify $RA$ and $\overline{RA}$ nodes for 8 out of 12 SF patients, and that, using the AUC of this classification (distinguishability $D'$) as a patient score, they could classify SF and NSF patients with an AUC of 0.76. In our study, however, we have only found an AUC of 0.65 when implementing their methodology, and an optimal value of AUC = 0.68 for the $\alpha_1$ band with the combined distinguishability score. These results are in agreement with those found in the main part of our study and with our previous findings with this same dataset.[46] Further studies are needed to elucidate the origin of the lower performance found here compared to Ramaraju *et al.*[48] We identify methodological considerations, such as the choice of connectivity metric—we considered here a phase metric, the PLI, that is insensitive to volume conduction, whereas Ramaraju *et al.*[48] used uncorrected amplitude correlations—or the thresholding procedure used (simple thresholding versus the disparity filter considered here). Moreover, the small dataset considered by Ramaraju *et al.*[48] could have driven the higher performance of the classification analysis. The findings may also reflect intrinsic differences between the patient populations: the cohort in this study is highly heterogeneous, including patients with different etiologies.

## Multi-frequency analysis

In our study, we considered a multi-band description, in analogy with some previous studies.[39,51,80,86-89] These studies found for the most part comparable results across frequency bands, with significant differences in brain network organization between epilepsy patients and controls, or between SF and NSF epilepsy-surgery patients, arising predominantly in the $\theta$ and $\alpha$ bands,[39,51,87-89] although differences have also been observed in the $\delta$ and $\gamma$ bands[86] and in the ripple and fast ripple bands.[86,87]

In our study, we also found comparable results across frequency bands for the node-based analyses, both at the individual and group level. Some metrics, such as the local clustering $c$, $\beta_0$ and $\beta_2$, however, only picked up differences between node sets in the broadband network. Notably, only in this band were the sizes of the $N$ and O node groups markedly different (when considering all ROIs and patients, see Supplementary Fig. 1 for more details). Regarding the patient classification analysis, we found the best classification for the broadband and $\alpha_1$ bands in agreement with the literature.[39,51,87,89] Remarkably, we found the strongest variations across frequency bands in the patient classification analysis (Fig. 5C, see also Supplementary Fig. 8 for details of the comparisons). Whereas in the broadband and the lower frequency bands (in particular $\delta$ and $\alpha_1$) we found a somewhat better outcome for patients with a high distinguishability score, this was not the case for higher frequency bands (in particular $\beta$ and $\gamma$, see Fig. 5). Overall, these results suggest that the hub-distribution changes differently across frequency bands for patients with good and bad outcomes. This finding highlights the need for further multi-frequency studies of epilepsy surgery to explore this effect in depth and determine its potential role in improving epilepsy-surgery planning.

## Methodological considerations

In this study, we considered the same patient database as in our previous study.[46] In this previous study, a machine learning analysis was used to classify network nodes as belonging or not to the RA, and to classify patients as having good (SF) or bad (NSF) outcomes. The performance of the node classifier was fair (60.37% accuracy), but the patient classification failed (49.03% accuracy). We have introduced several methodological changes relative to this original study, from the consideration of multiple frequency bands, the three-node-group partition, and the inclusion of node-neighbourhoods and topological data analysis. The methodologies of the two studies can be compared via the betweenness centrality, a benchmark centrality measure considered in both studies: Nissen *et al.*[46] found that hub nodes overlapped more than expected by chance with the $RA$. This is in qualitative agreement with our finding that RA nodes are, at the group level, significantly more central than $\overline{RA}$ nodes.

Regarding patient classification, Nissen *et al.*[46] performed a classification based on a combination of individual and average metrics, namely the averages over (i) $RA$ nodes, (ii) the resection lobe, (iii) nodes contralateral to the $RA$, (iv) $\overline{RA}$ nodes, and two metrics measuring the difference between the average over $RA$ and the contralateral nodes, and over $RA$ and $\overline{RA}$ nodes. No significant differences between SF and NSF patients were identified at the group level, and a machine learning analysis was also unable to classify the patients according to surgical outcome. In our study, instead of using the centrality values directly, we exploited the results of the node-based analyses to perform a patient classification analysis, similarly to the analysis by Ramaraju *et al.*[48] In particular, we defined a distinguishability score based on the difference between each of the node sets, and we found an AUC of 0.68 for the broadband network (the same as used by Nissen *et al.*[46]). In this manner we were able to exploit a patient-specific analysis, accounting for heterogeneity in the patient population, which can be lost if comparisons of absolute values among patients are performed. The differences in findings between the two studies, and our finding that a population-based analysis is less sensitive than the patient-specific analysis, highlight the need to consider methodologies that allow for individualized patient characterization.[23,24]

Whereas some of the studies mentioned above,[48] as well as other recent studies,[23] have found better classification results than the ones found in this study, the strength of this study lies in the much larger patient cohort considered here, which is two to three times larger than typical cohort sizes in similar studies. Moreover, we further validated the robustness of our findings with respect to several methodological choices, including the frequency band of the MEG-based brain networks and specific analysis details, benchmarking our findings and analysis pipelines against previous studies.[46,48]

# Conclusion

Pathological hubs occur in the brain networks of refractory epilepsy patients that do not necessarily overlap with the EZ, but may instead be strongly connected to it. Thus, a positive surgical outcome may also be obtained if the surgical resection does not include a pathological hub. In this study, we have found that a three-group partition of the brain regions, where the neighbours of the RA are separated from the remaining brain regions, can provide novel information regarding the organization of the epileptogenic network. Regional descriptors of hub status and network organization, as the ones we propose here based on the notion of extended neighbourhoods, provide new tools to characterize the effect of a proposed resection. Our findings also evidence the heterogeneity of the patient population, and the need for individualized studies that allow for a patient-specific consideration of brain connectivity.

# Supplementary material

Supplementary material is available at *Brain Communications* online.

# Funding

Funding for open access charge: Universidad de Granada / CBUA. A.P.M. acknowledges financial support by the 'Ramón y Cajal' program of the Spanish Ministry of Science and Innovation (grant RYC2021-031241-I) and from the Spanish State Research Agency, Project PID2020-113681GBI00, financed by MICIN/AEI/10.13039/501100011033. A.P.M. and I.N. were supported by ZonMw and the Dutch Epilepsy Foundation, project number 95105006, and Dutch Epilepsy Foundation project 14-16 (I.N.) L.D.G. acknowledges financial support from the Clinical Neurophysiology Department (KNF) at the Amsterdam University Medical Center (AUMC). N.D. acknowledges funding by the Swiss National Science Foundation (SNSF) under project funding ID: 200021 207537 and by the Deutsche Forschungsgemeinschaft (DFG, German Research Foundation) under Germany's Excellence Strategy EXC2181/1-390900948 (the Heidelberg STRUCTURES Excellence Cluster). This work was partially supported by a grant from the Simons Foundation (G.B.). G.B. and A.P.M. would like to thank the Isaac Newton Institute for Mathematical Sciences, Cambridge, for support and hospitality during the programme *Hypergraphs: Theory and Applications*, where work on this paper was undertaken. This work was supported by grant EP/R014604/1 of the Engineering and Physical Sciences Research Council (EPSRC). The funding sources had no role in study design, data collection and analysis, interpretation of results, decision to publish, or preparation of the manuscript.

# Competing interests

The authors declare that they have no competing interests.

# Data availability

The data used for this manuscript are not publicly available because the patients did not consent for the sharing of their clinically obtained data. Requests for access to the data should be directed to the corresponding author. All user-developed codes are available on GitHub: https://github.com/LeonardoDiGaetano/TDA-Epilepsy.